\documentclass[11pt,twoside]{article}
\usepackage{./asp2014}
\usepackage{color}

\newcommand{\Msun}{$M_{\odot}$}

\newcommand{\uJybm}{$\mu$Jy\,beam$^{-1}$}
\def\farcs{''\mskip-9mu.\,}    

\hypersetup{
 colorlinks= true, 
 urlcolor  = black, 
 linkcolor = red, 
 citecolor = blue 
} 

\aspSuppressVolSlug
\resetcounters

\begin{document}

\title{Probing Obscured MBH Accretion and Growth since Cosmic Dawn}
\author{Wiphu Rujopakarn$^{1}$, Kristina Nyland$^2$, and Amy E. Kimball$^3$\\
\affil{$^1$Department of Physics, Faculty of Science, Chulalongkorn University, 254 Phayathai Road, Pathumwan, Bangkok 10330, Thailand; \email{wiphu.r@chula.ac.th}}
\affil{$^2$National Radio Astronomy Observatory, Charlottesville, VA, 22903, USA; \email{knyland@nrao.edu}}
\affil{$^3$National Radio Astronomy Observatory, Socorro, NM, 87801, USA; \email{akimball@nrao.edu}}
}

\paperauthor{Wiphu Rujopakarn}{wiphu.r@chula.ac.th}{}{Chulalongkorn University}{}{Pathumwan}{bangkok}{10330}{Thailand}
\paperauthor{Kristina Nyland}{knyland@nrao.edu}{}{National Radio Astronomy Observatory}{}{Charlottesville}{VA}{22903}{USA}
\paperauthor{Amy Kimball}{akimball@nrao.edu}{}{National Radio Astronomy Observatory}{}{Socorro}{NM}{87801}{USA}

\begin{abstract}
Most of the stars today reside in galactic spheroids, whose properties are tightly tied to the supermassive black holes (MBHs) at their centers, implying that the accretion activity onto MBHs leaves a lasting imprint on the evolution of their host galaxies.  Despite the importance of this so-called MBH-galaxy co-evolution, the physical mechanisms responsible for driving this relationship -- such as the dominant mode of energetic feedback from active galactic nuclei (AGN) -- 
remain a poorly understand aspect of galaxy assembly.  A key challenge for identifying and characterizing AGN during the peak epoch of galaxy assembly and beyond is the presence of large columns of gas and dust, which fuels the growth of their MBHs but effectively obscures them from view in optical and X-ray studies.  The high sensitivity of the ngVLA will capture emission from AGN in an extinction-free manner out to $z \sim 6$ and beyond.  At lower-redshifts ($z \sim 2$), the high angular resolution of the ngVLA will enable spatially-resolved studies capable of localizing the sites of actively growing MBHs within their host galaxies during the peak epoch of cosmic assembly.  
\end{abstract}

\section{Introduction}
The convergence of theory and observation in galaxy evolution is a major milestone in astrophysics. With appropriate sets of parameters, models of galaxy formation can reproduce, for instance, the luminosity and mass distributions of today's galaxies and the cosmic star formation rate (SFR) history since the epoch of reionization.  However, many of the most fundamental processes in the model are not well understood, especially down to galactic scales, where the current frontier questions in galaxy evolution lie.  Central among these issues is the empirical link between galaxy assembly and supermassive black hole (MBH) accretion and growth. Advancing our understanding of this symbiotic relationship requires a complete census of the first generation of AGN at cosmic dawn as well as a spatially-resolved study of individual galaxies at the peak of their assembly, $1 < z < 3$, to map their star formation (SF) and MBH accretion activities (i.e., active galactic nuclei; AGN).  Fundamental breakthroughs in these areas thus require a sub-arcsecond resolution, extinction-independent tracer of {\it both} AGN and SF.   

Centimetric radio continuum observations uniquely meet these observational demands. Deep, high-resolution continuum surveys with current and next-generation facilities will therefore play an essential role in advancing the current galaxy formation and evolution paradigm in the decades to come.  In addition to tracing AGN and SF activity in an extinction-free manner, centimetric radio continuum observations of AGN and SF emission are also relatively stable in time. While other AGN indicators such as hard X-ray emission or emission lines can fluctuate rapidly over timescales of days to months \citep{Hickox14}, giving only an instantaneous indication of the level of activity, radio emission has a synchrotron timescale of $10^6 - 10^7$ years \citep{BeckKrause05} to serve as a beacon localizing the AGN. Obviously, a prerequisite to utilizing these characteristics is to confirm the AGN origin of the radio emission, which, by nature, traces both AGN and SF activities. Thanks to the Atacama Large Millimeter/submillimeter Array (ALMA), this is now possible out to $z \sim 10$ with the help of the negative $K$-correction. As a result, we are reaching the start of an era when centimetric radio continuum can be utilized to the fullest extent to study MBH growth. 

\section{Current Observational Prospects}
\subsection{Identifying Obscured AGN at $z \gtrsim 6$}

Luminous AGN have been found out to $z = 7.5$ with $M_{\rm BH} \sim 10^9$ \Msun\ \citep{Banados18}. Such extreme objects are rare (only two with spectroscopic confirmation at $z > 7$ so far), yet provide crucial evidence that MBH can grow rapidly up to massive size within a Gyr after the Big Bang. If the Ba{\~n}ados et al. quasar is at the bright-end of the quasar luminosity function, it remains a question whether more typical MBHs are growing rapidly elsewhere during this epoch. Typical, actively growing MBHs at $z > 6$ appear to be beyond the present observational capabilities: none of the spectroscopically confirmed $z = 6-8$ galaxies in the Hubble Ultra-Deep Field (HUDF) are detected in the individual or stacked {\it Chandra} 4 and 7 Ms images \citep{Giallongo15, Luo17}. \citet{Treister13} postulate that these galaxies may not contain MBH, that MBH are not growing despite their large sSFR, or that MBHs are growing elsewhere in lower-mass or obscured galaxies not identified by the current {\it Hubble} and {\it Chandra} observations. This missing typical population of AGN could have important implications on the cosmic accretion history, feedback on the intergalactic medium, reionization, and to our understanding of how MBHs and galactic bulges coevolved.

\begin{figure}[t!]
\begin{center}
\includegraphics[width=0.88\textwidth]{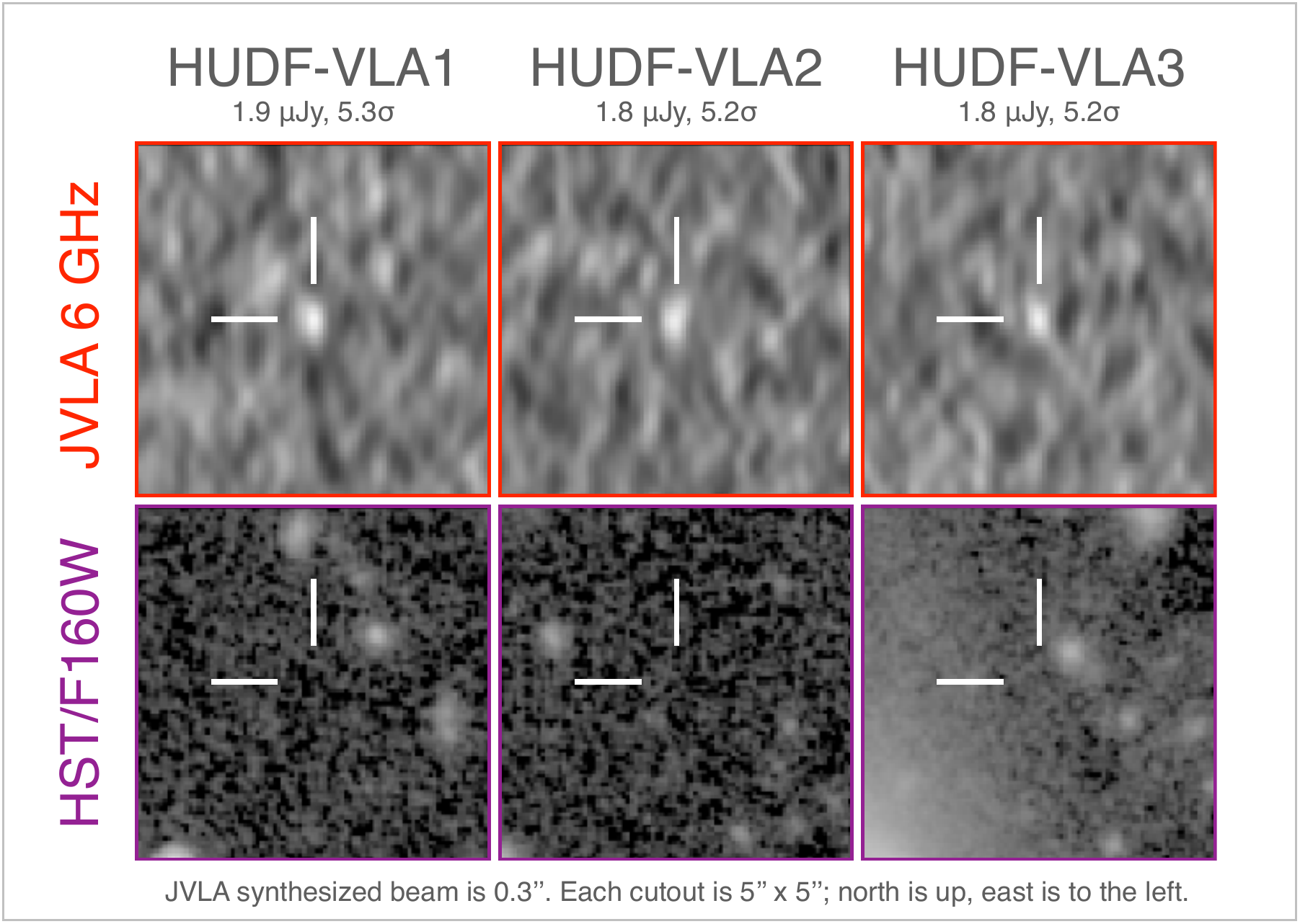}
\caption{The $\geqslant5\sigma$ sources in the HUDF JVLA 6 GHz survey with absolutely no counterparts in any other bands ({\it Hubble} 1.6 \micron\ shown here). An object has to be at $z \gtrsim 5$ in order to evade detection in ultra-deep {\it Hubble, Spitzer}, and ground-based imaging sensitivity of the HUDF. There are no sources in the negative map at this significance --- it is extremely unlikely that these are noise peaks.}
\end{center}
\vspace{-10pt}
\end{figure}

If the typical sites of MBH growth in the early universe are indeed in low-mass galaxies, the current optical, IR, and X-ray indicators will have difficulties identifying them. They will be too faint in the optical/near-IR, contain too little dust to power mid-IR emission, and emit at too low of a luminosity for direct X-ray detection at high redshift (also not identified in large numbers in other bands to enable X-ray stacking). On the other hand, early MBHs could be accreting in a different mode that confers on them the observed rapid growth \citep[e.g.,][]{VolonteriSilkDubus15}, but that the radiation is trapped \citep{CoughlinBegelman14} and conceals them from current survey approaches. In either case, centimetric radio observations have the potential to identify them because of the symbiotic nature of the accretion onto MBH and the resulting relativistic jets that are a signpost of AGN in the radio \citep[e.g.,][]{BegelmanBlandfordRees84}. Radio observations are also extinction-free and have recently become sensitive enough to find typical AGN at $z \gg 4$ with the advent of the JVLA. This is therefore a potentially groundbreaking means to identify typical AGN at high $z$.

Recently, \citet{Rujopakarn16} has conducted a sensitive single-pointing survey of the HUDF using the JVLA at 6 GHz (0.3 \uJybm\ rms). They have found that 99\% of radio sources detected above 5$\sigma$ have optical or near-IR counterparts. The remaining 1\% -- three sources within the central $100''$ of the primary beam -- are of great interest (Figure 1). They are extremely unlikely to be noise peaks because the noise distribution from the source-free area is well-fitted by a Gaussian and there are no sources in the inverted map above $5\sigma$ in the same area. They are also persistent over the two year period to acquire 177 hours of JVLA observations, effectively ruling out the possibilities of transient or non-astrophysical artifacts. To evade detection in a field with such a deep suite of ground and space-based imaging, reaching 30.5 mag ($5\sigma$) at 1 \micron, an object must either be at moderate redshifts, $z \sim 1-3$, but extremely dust-obscured, which is ruled out by the lack of detections with {\it Spitzer} and ALMA; or have low $M_*$ (e.g., $10^7-10^8$ $M_{\odot}$, which will be an interesting class of AGN in itself); or they must be at higher $z$ of $z \gtrsim 5-6$. 

\begin{figure}[t!]
\begin{center}
\includegraphics[width=\textwidth]{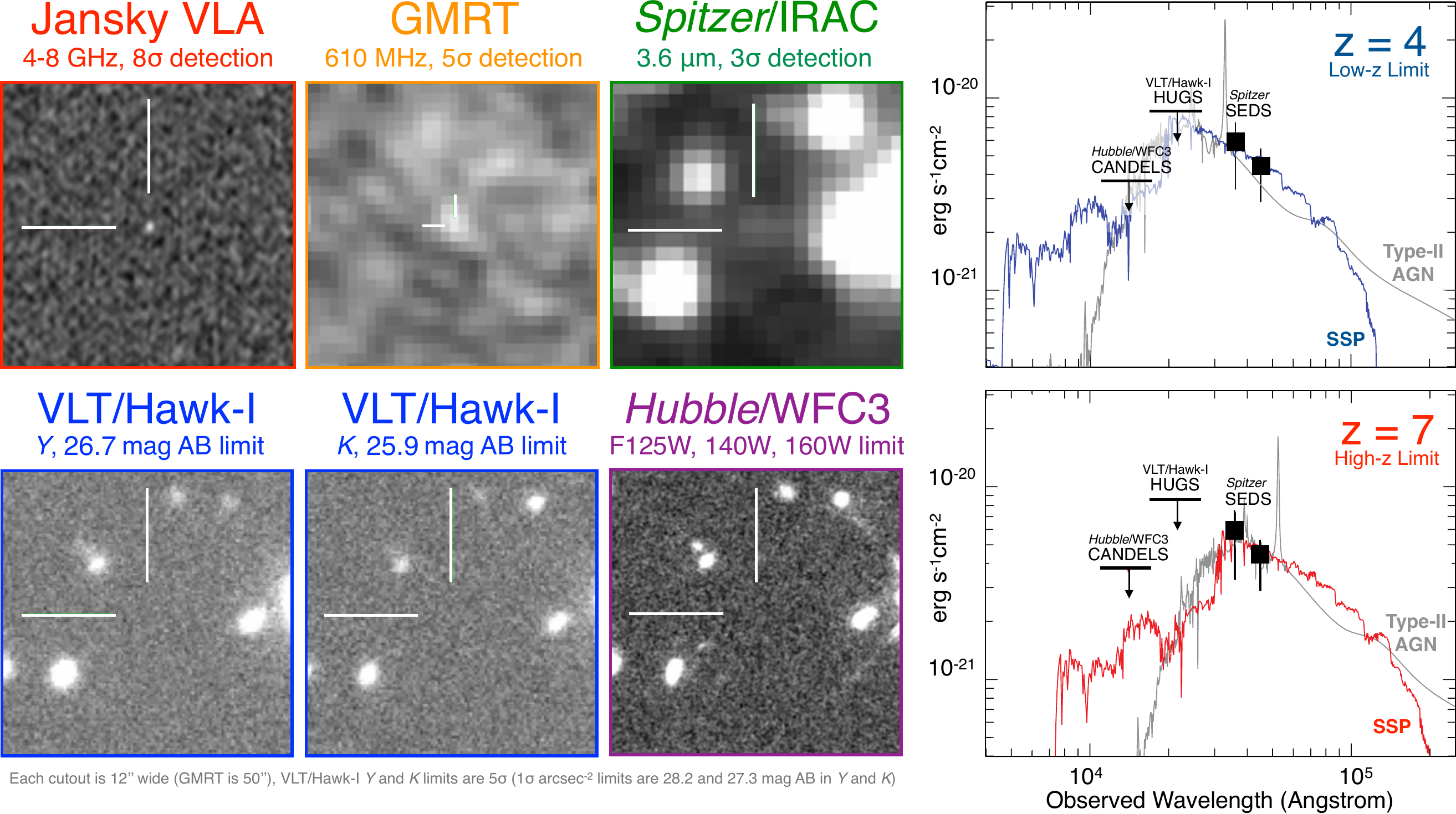}
\caption{A high-$z$ AGN candidate detected in the pilot survey of \citet{Rujopakarn16} that is only detected in the VLA and an ultra-deep follow-up imaging with the GMRT and {\it Spitzer}/IRAC, but not any other bands; the most probable redshift is $z \sim 4-7$. Follow-up spectroscopy with the ESO VLT/X-Shooter with 6-hours integration finds no Ly$\alpha$, typical for $z > 6$ candidates, highlighting the necessity for next-generation facilities, e.g., {\it JWST}, to study these populations.}
\end{center}
\vspace{-10pt}
\end{figure}

Characterization of these high-$z$ radio AGN candidates has proven to be extremely challenging for current facilities. An example is a candidate shown in Figure 2 from the pilot survey of \citet{Rujopakarn16}. In this case, a faint {\it Spitzer}/IRAC counterpart is found, but the non-detection in deep {\it Hubble} and ground-based near-IR images effectively ruled out hosts with $M_* \gtrsim 10^8$ $M_{\odot}$ at $z \lesssim 4$. This radio source was followed-up with the X-Shooter instrument (simultaneous UV, optical, and near-IR spectroscopy) on the ESO Very Large Telescope for 6 hours to search for a Ly$\alpha$ line to confirm the redshift. No line was detected, suggesting that this could be a case of the low Ly$\alpha$ detection rate at $z \gtrsim 6.5$, where the reionization is not yet completed \citep{Schenker14}. Their characterization is beyond the capability of current optical/IR observatories and will require next-generation facilities, e.g., {\it JWST}.

\begin{figure}[t]
\begin{center}
\includegraphics[width=\textwidth]{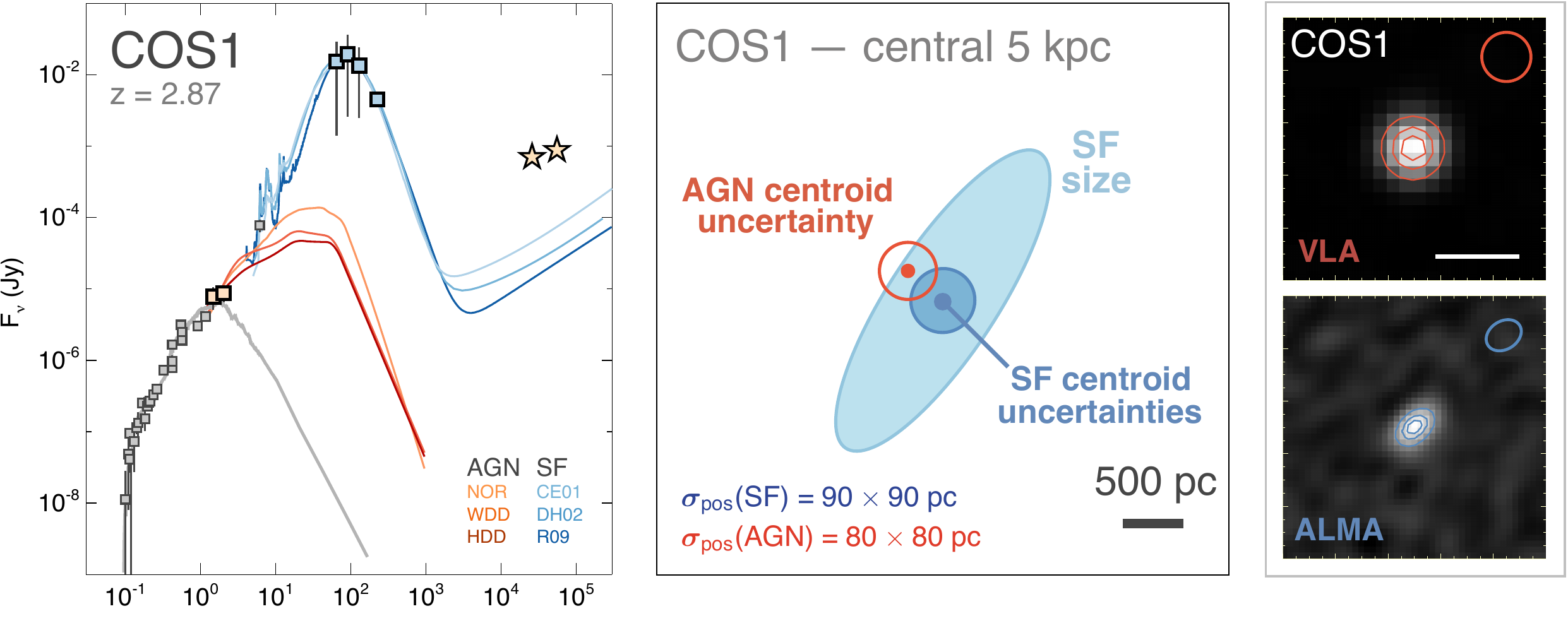}
\caption{A new technique to accurately pinpoint the sites of intense SF and MBH growth at $z \sim 2$. Left: strong radio excess over the level associated with SFR is AGN-dominated, whereas the submillimeter emission captured by ALMA is virtually free of the AGN emission and hence trace the distribution of SF. Middle: Using this technique, \citet{Rujopakarn18} found AGN to occur within the compact regions of intense SF. Right: Accurate localization of SF and AGN are enabled by sub-arcsec ALMA and JVLA observations. Postage stamps are $4'' \times 4''$; the synthesized beam size is shown in the respective image.}
\end{center}
\vspace{-10pt}
\end{figure}

\subsection{MBH Localization at $z \sim 2$}
\label{sec:localization}
Despite the multiple lines of evidence showing a link between galaxy assembly and MBH growth, there is no consensus on the causality of this relationship. Direct links on a global scale between a MBH accretion rate (BHAR) and the star formation rate (SFR) of its host are elusive. Although the two processes appear to be correlated, this result may only reflect that both depend on the stellar mass of the host galaxy \citep{XuRieke15, Stanley17}. For local galaxies, there is evidence for {\it localized} enhanced SF close to an AGN, related to the BHAR \citep[e.g.,][]{DiamondStanic12}. Improving our understanding of this link at higher redshifts requires higher resolution imaging to determine whether the sites of MBHs are also associated with the most vigorous SF.  Yet, such images have until very recently been impossible at $z \sim 2$, at the peak of both SFR and BHAR and thus potentially the formative era for the current relation.

The JVLA provides sub-arcsecond imaging that can penetrate dust to trace synchrotron emission associated with SF as well as emission from any AGN core and jets. The submillimeter continuum at, e.g., 870 \micron\ traces the cold dust emission associated with SF. This emission is virtually free of AGN contribution because the IR spectra of AGN tori plummet rapidly longward of 40 \micron\ \citep{LyuRieke17}. Likewise, extrapolating the radio slopes from $\sim$1 GHz to the ALMA band shows that the non-thermal emission from all but the most radio-loud AGN is likely to be $2-3$ orders of magnitude fainter than the thermal dust emission of their hosts.  Any galaxy where the radio emission is enhanced to well above the level implied by the far-IR/radio correlation for SF galaxies presents a robust radio AGN signature (Figure 3). In these galaxies with strong radio enhancement, JVLA traces AGN and ALMA traces nearly pure SF, so sub-arcsecond JVLA and ALMA images can localize --- at sub-kpc resolution --- the sites of SF relative to the MBH in active galaxies.

Central to this technique is that (1) at $z \sim 2$, AGN commonly have strong radio sources, unresolved at sub-arcsec resolution \citep[cf. parsec-scale cores and jets;][]{Zensus97}; (2) at GHz rest frequencies, the central radio emission within a sub-arcsec beam accurately pinpoints the AGN core \citep{Nagar02, PushkarevKovalev12}; (3) the uncertainty of the centroid position, $\sigma_{\rm pos}$, of a radio or submillimeter source is $\sigma_{\rm pos} \approx \theta_{\rm beam}/(2 \times {\rm S/N})$, where $\theta_{\rm beam}$ is the synthesized beam size and S/N is the peak signal-to-noise ratio \citep{Condon97}. For example, a 10$\sigma$ point source observed with a $0\farcs15$ beam can be localized to about 8 milliarcsecond (mas); and, lastly (4) the astrometric references of the JVLA and ALMA are tied to their respective phase calibrator positions, which are in turn tied to the International Coordinate Reference Frame (ICRF) to within $\lesssim10$ mas. That is, the astrometry of JVLA and ALMA are tied to the common reference frame and can be directly compared.

\citet{Rujopakarn18} demonstrated this technique in three objects with $\geqslant5\times$ radio excess over the level predicted by the far-IR/radio correlation, and with the ALMA observations resolving the sub/millimeter emission. In all three galaxies, the AGN are located within the $\lesssim1$ kpc regions of gas-rich, heavily obscured, intense nuclear SF (Figure 3). If the star-formation and AGN-driven outflows that may be present \citep[e.g.,][]{Mullaney13} do not completely disrupt the cold gas supply, and both types of activity proceed for the duration of their host's typical stellar mass doubling time of $\simeq 0.2$ Gyr, then the newly formed stellar mass within the central kpc will be of order $10^{11}$ \Msun. Likewise, if the MBH is allowed to accrete at a rate predicted by the correlation between SFR and the average BHAR relation \citep[e.g.,][]{Chen13} for the same duration, the accreted MBH mass will be $10^{6.7} - 10^{7.8}$ \Msun, similar to those found in local massive galactic bulges. This is consistent with {\it in situ} galactic bulge and MBH growth, and may represent the dominant process regulating the bulge--MBH relationship through which all massive galaxies may pass.

\section{Anticipated Results with the ngVLA}
\subsection{An All-Sky Survey with the ngVLA}
Although the JVLA has proven to be an excellent tool for the detection of high-redshift AGN, such observations are time consuming, particularly in the case of all sky surveys.  A survey with the ngVLA carried out in a similar manner to the VLA Sky Survey (VLASS) would offer improvements in efficiency due to the wider field of view of the 18m dishes and higher collecting area compared to the VLA.  An all-sky survey with the same total observing time as VLASS ($\approx$6000~hours) could be $\approx$10 times deeper ($7 \,\mu$Jy~beam$^{-1}$), thereby capturing high-redshift AGN populations only detectable in the radio.  Such a survey would be expected to identify $\sim$ 10$^7$ AGN at sub-arcsecond angular resolution \citep{Nyland18}.

\subsection{Ultra-deep ngVLA Square Degree Survey}
The AGN localization study with the JVLA described in \citet{Rujopakarn18} offered an important demonstration of the need for both high resolution and depth for pinpointing the sites of actively growing MBHs.  However, the \citet{Rujopakarn18} study covered only a single JVLA pointing at C-band, and required 177 hours of integration time.  The ngVLA will allow this technique to be applied to much larger sky areas and offer an even higher degree of positional accuracy.  The milliarcsecond resolution of ngVLA will push the localization accuracy of the AGN to the sub-milliarcsecond regime using the technique described in Section~\ref{sec:localization}.  Most importantly, it will resolve the morphologies of these sub-kpc jets at $z \sim 2$ to understand their role in regulating the assembly of the bulge \citep{Nyland18}.  Because there are no local counterparts to gas-rich bulge-forming galaxies, ngVLA will be the only avenue for providing theorists and simulators with constraints on the nature of typical AGN-driven feedback at the peak of galaxy assembly.  

An ngVLA 8~GHz survey over 1 deg$^2$ with a 1$\sigma$ depth of 300 nJy~beam$^{-1}$ and an angular resolution of 0.1$^{\prime \prime}$ would provide an ideal combination of area, depth, and resolution to robustly pinpoint the relative locations of galactic bulge and MBH growth.  Extrapolating from the current ngVLA reference design, such a survey would require $\approx$ 200 hours.  We note thast this survey could be conducted as a high-resolution component to that described in the Barger et al. chapter in this volume, and would also provide as an added bonus information on extinction-free morphologies of SF.

\bibliography{ngVLA_accretion_cosmic_dawn_v3}  

\end{document}